\documentclass[10pt]{article}

\usepackage{amssymb}
\usepackage{amsmath}
\usepackage{amsthm} 
\usepackage{mathrsfs}                                                           
\usepackage{slashbox}  
\usepackage[all]{xy}
\usepackage{graphicx}
\usepackage{wrapfig}

\def\2{{1\over 2}}

\def\d{{\mathrm{d}}}
\newcommand{\rf}[1]{(\ref{#1})}
\def\b{\bar}
\newcommand{\ud}{\mathrm{d}}
\renewcommand{\t}{\tilde}
\newcommand{\p}{\partial}

\newcommand{\mA}{\mathbf{A} }
\newcommand{\mB}{\mathbf{B} }
\newcommand{\mC}{\mathbf{C} }
\newcommand{\mD}{\mathbf{D} }
\newcommand{\mW}{\mathbf{W} }
\newcommand{\mU}{\mathbf{U} }

\newcommand{\mF}{\mathcal{F} }
\newcommand{\mH}{\mathcal{H} }
\newcommand{\mG}{\mathcal{G} }
\newcommand{\mQ}{\mathcal{Q} }
\newcommand{\mK}{\mathcal{K} }
\newcommand{\m}{\mathfrak{m}}
\newcommand{\mg}{\mathfrak{g}}
\newcommand{\di}{\mathbf{div}}
\newcommand{\mb}{\mathbf{b}}
\renewcommand{\mD}{\mathcal{D}}

\title{
\bf{SFT-inspired Algebraic Structures in Gauge Theories}}
\author{Anton M. Zeitlin
\footnote{anton.zeitlin@yale.edu http://math.yale.edu/$\sim$az84 http://www.ipme.ru/zam.html}   \\
Department of Mathematics,\\
Yale University,\\
442 Dunham Lab, 10 Hillhouse Ave\\
New Haven, CT 06511}
\date{}

\begin{document}
\maketitle
\begin{abstract}
We consider gauge theories in a String Field Theory-inspired formalism. The constructed algebraic 
operations lead in particular to homotopy algebras of the related BV theories. We discuss  invariant description of the gauge fixing procedure and special algebraic features of gauge theories 
coupled to matter fields.
\end{abstract}
\section{Introduction}
The $L_{\infty}$ algebras (or homotopy Lie algebras) \cite{stasheff} first entered physics in the context of 
study of higher-spin particles \cite{highspin}. Soon after that, the same structures appeared in the mathematical treatment \cite{sta} of the Batalin-Fradkin-Vilkovisky 
approach \cite{bv} (see also \cite{schbv}, \cite{stasheff2}). 
In the beginning of the 1990s, these generalizations of Lie algebras appeared in 
Zwiebach's formulation of the closed String Field Theory (SFT) \cite{zwiebach}.  
Namely, the homotopy Lie algebra was the algebra of gauge 
symmetries of the theory. The action for closed SFT was given by the homotopic generalization of the Chern-Simons 
theory \cite{schwarz}-\cite{krotovlosev}, and was represented as 
\cite{zwiebach}:
\begin{eqnarray}\label{csint}
S=\frac{1}{2}\langle\Psi, Q\Psi \rangle+\sum^{\infty}_{n=3}\frac{\kappa^{n-2}}{n!}\{\Psi,...,\Psi\},
\end{eqnarray} 
where $\Psi$ is a string field, $\{\cdot,...,\cdot\}=\langle\cdot, [\cdot,....,\cdot]\rangle$ and $[\cdot,...,\cdot]$ generate, together 
with the nilpotent operator $Q$, the $L_{\infty}$ algebra, and $\langle\cdot, \cdot\rangle$ is an inner product. 
The equation of 
motion for such an action is the Generalized Maurer-Cartan equation associated with this homotopy Lie algebra:
\begin{eqnarray}
Q\Psi+\sum^{\infty}_{n=2}\frac{\kappa^{n-1}}{n!}[\Psi,...,\Psi]=0.
\end{eqnarray}
A similar action appeared in open SFT \cite{witten}-\cite{taylor1}, but there, the algebraic structure was based on the  $A_{\infty}$ algebra. 

In \cite{zeit3}, we considered formal Maurer-Cartan structures which appear naturally in the study of 
the deformation of the string theory BRST operator \cite{brst} in the case of conformal field theory  corresponding to open and closed strings. In the open string case, we were able to build the bilinear operation acting on string theory operators, which has a form of regularized commutator 
\begin{eqnarray}
R(\phi^{(0)},\psi^{(0)})(t)=\mathcal{P}[\phi^{(0)}(t+\epsilon),\psi^{(0)}(t)]-(-1)^{n_\phi n_{\psi}}
\mathcal{P}[\psi^{(0)}(t+\epsilon),\phi^{(0)}(t)],
\end{eqnarray}
(here $\mathcal{P}$ stands for the projection on the $\epsilon$-independent part, $\phi^{(0)},\psi^{(0)}$ are vertex operators on the boundary of the worldsheet and $n_\phi$, $n_{\psi}$ represent their ghost numbers) such that the resulting generalized Maurer-Cartan equation 
\begin{eqnarray}\label{gmc}
[Q_{BRST},\phi^{(0)}]+ \frac{1}{2}R(\phi^{(0)},\phi^{(0)})+...=0
\end{eqnarray}
(where $Q_{BRST}$ is the usual BRST operator for open string) 
for appropriate choice of operator $\phi^{(0)}$ of ghost number equal to 1, leads to Yang-Mills equations up to the second order. Symmetries of the equation \rf{gmc}
\begin{eqnarray}
\phi^{(0)}\to \phi^{(0)}+\epsilon([Q_{BRST},\lambda^{(0)}] +R(\phi^{(0)},\lambda^{(0)})+...),
\end{eqnarray}
where $\lambda^{(0)}$ is some operator of ghost number 0, 
reproduce the gauge symmetries of the YM system. However, it was not clear how to build the third order operation on the conformal field theory level. Nevertheless, it was possible to do the following. First, to 
shorten the BRST complex of open string, leaving only operators, corresponding to the light sector. 
The shortened BRST complex actually coincides with the following complex of differential forms:
\begin{eqnarray}\label{ymc}
0\to\mg\xrightarrow{i}\Omega_{\mg}^{0}\xrightarrow{\d}\Omega_{\mg}^{1}
\xrightarrow{*\d *\d}\Omega_{\mg}^{1}\xrightarrow{*\d *}\Omega_{\mg}^{0}\to 0,
\end{eqnarray} 
where $\Omega_{\mg}^{i}$ is the space of $\mg$-valued ($\mg$ is a reductive Lie algebra) differential 
forms of degree $i$, $d$ is a de Rham differential and $*$ is a Hodge star.  
After that, one can explicitly construct the third order operation on the resulting shortened complex.  
This was done in  \cite{ym}, \cite{bvym}. There we found that one can define this trilinear operation $[\cdot, \cdot,\cdot]_h$ on the shortened BRST complex $(\mF^{\cdot},Q)$ in such a way that  $(\mF^{\cdot},Q) $
together with $[\cdot, \cdot]_h, [\cdot, \cdot,\cdot]_h$ (where $[\cdot, \cdot]_h$ stands for operation $R$) forms an $L_{\infty}$ algebra.  The other possible $n$-linear operations are set to be equal to zero.
  
 This hidden nontrivial homotopy Lie algebra structure 
inside (nonabelian) Yang-Mills theory allowed us to express the Yang-Mills action and its BV extension in the form \rf{csint}. The 
gauge symmetries were shown to coincide with the symmetries of the corresponding Maurer-Cartan equation, which in turn 
coincides with the equation of motion in the Yang-Mills theory. 

 In \cite{bvym}, we 
emphasized the algebraic similarity between the ordinary 3-dimensional Chern-Simons theory and 
the Yang-Mills theory (this is due to the fact that the de Rham complex in three 
dimensions is as short as \rf{ymc}). 
In section 2, we quickly review the results of \cite{ym} and \cite{bvym} which led to the $L_{\infty}$ algebra of the pure gauge 
theory. 

The aim of this paper is to extend the complex \rf{ymc} and therefore its algebraic structures in order to include other 
fields which interact nonlinearly, but obviously in a gauge-invariant way with the nonabelian Yang-Mills theory. However, before 
that, in section 3, we introduce a construction, in addition to those considered in \cite{bvym}, which is necessary for the quantum theory. 
Namely, we discuss the gauge fixing procedure for the Yang-Mills action. We extend the complex \rf{ymc} by adding the spaces 
corresponding to the antighost field and its antifield. In this way we arrive at the elegant formula \rf{genfun} using the decomposition \rf{decomp}, which describes on an algebraic level the Yang-Mills theory in the Lorenz gauge. The construction is closely related to what is known as the Siegel gauge in open SFT \cite{pol}. 
Moreover, in subsection 3.3, we give some nontrivial extensions of 
the $L_{\infty}$ algebra on this extended complex, using the data of 
bilinear operation from \cite{zeit3}  based on the operator product expansion. This homotopy algebra turns out to be very unusual and strange from the field theory point of view, since it has no invariant inner product unlike $L_{\infty}$ algebras related to the BV formalism. Therefore, the field theory interpretation of this extension is not yet clear.

Sections 4 and 5 are parallel: we consider the homotopy algebras related to the Yang-Mills theory with matter (scalar and spinor) fields.

In section 4, we consider our first example of matter fields entering the theory. Namely, we consider the scalar field which 
takes values in the adjoint representation of $\mg$ interacting with the Yang-Mills field. We show that one can generalize 
the bilinear and trilinear operations in such a way that it allows to express this theory in the same homotopy 
Chern-Simons form. 

Section 5 is devoted to the Dirac fermion field interacting with a gauge field. This theory turns out to be poorer in the algebraic sense. In order to describe its homotopic structure, one just needs to generalize only 
the bilinear operation, but the 3-linear one is set to 
be equal to zero when one of the arguments belongs to the extension of the complex \rf{ymc}. It is possible to gues that from 
the nonlinear structure of the interaction term (it is cubic: quadratic w.r.t. the fermion field and linear w.r.t. the gauge field).

It should be noted, that  the local version (when all the fields are constant) of 
$L_{\infty}$ algebras, corresponding to the Yang-Mills theory 
was considered in \cite{movshev}. 

One should also note that since the original paper \cite{schkonts}, where the abstract statement about the relation between the BV formalism 
and $L_{\infty}$ algebras was given, there has been a lack of consideration of explicit field theory examples. Namely, except for \cite{movshev} the explicit 
examples were different versions of the Chern-Simons theory only, where the algebraic structure is given by the simple graded differential Lie algebra, which is a very particular case of $L_{\infty}$ algebra. We hope that here we partly fill this gap.

In the last section, we bring up some open questions 
and discuss possible ways to give a homotopy  description of Gravity.

\section{The $L_{\infty}$ algebra and the (BV) Yang-Mills theory}

In this section, we specify the notation and give a short overview of the results obtained for the pure Yang-Mills case in \cite{ym}, \cite{bvym}. 

\vspace{3mm}
  
\noindent {\bf 2.0. Notations.} 
In this paper, we  will encounter two bilinear operations $[\cdot,\cdot]$, $[\cdot,\cdot]_h$. The first one,
without the subscript,  denotes 
the Lie bracket in the given finite-dimensional Lie algebra $\mathfrak{g}$. The second one,
 with subscript $h$, denotes the graded antisymmetric bilinear operation 
in the homotopy Lie superalgebra. 

We will use three types of operators 
acting on differential forms with values in some finite-dimensional reductive Lie algebra 
$\mg$. 
The first one is the de Rham operator $\ud$. The second one is the {\it Maxwell operator} 
$\mathfrak{m}$, which maps 1-forms to 1-forms. If $\mA=A_{\mu}dx^{\mu}$ is a 1-form, then 
\begin{eqnarray}
\m\mA\equiv*\ud*\ud\mA=(\p_{\mu}\p^{\mu}A_{\nu}-\p_{\nu}\p_{\mu}A^{\mu})dx^{\nu}, 
\end{eqnarray}
where indices are raised and lowered with respect to the metric $\eta^{\mu\nu}$. 
The third operator maps 1-forms to 0-forms, this 
is the operator of divergence $\di$. For a given 1-form $\mA$, 
\begin{eqnarray}
\di\mA\equiv*\ud*\mA=\p_{\mu}A^{\mu}. 
\end{eqnarray}
For $\mg$-valued 1-forms, one can also define the following (anti)symmetric bilinear and trilinear operations:
\begin{eqnarray}
&&(\mA,\mB)\equiv *(\mA,*\mB)_K=(A_{\mu},B^{\mu})_K,\nonumber\\
&&\{\mathbf{A},\mB\}\equiv*[\mB,*\ud\mA]+*[\mA,*\ud\mB]+*\ud *[\mA, \mB]\\
&&\mA\cdot\mW\equiv *[\mA,*\mW],\nonumber\\
&&\{\mA,\mB,\mC\}=*[\mA,*[\mB,\mC]]+*[\mC,*[\mA,\mB]]+*[\mB,*[\mC,\mA]],
\end{eqnarray}
where $(\cdot,\cdot)_K$ is the canonical invariant form on the Lie algebra $\mg$.

\vspace{3mm}

\noindent{\bf 2.1. The Yang-Mills chain complex.} 
Now we will give the explicit realization of the chain complex 
\begin{eqnarray}\label{cmplxforms}
0\to\mg\xrightarrow{i}\Omega_{\mg}^{0}\xrightarrow{\d}\Omega_{\mg}^{1}
\xrightarrow{\m}\Omega_{\mg}^{1}\xrightarrow{\di}\Omega_{\mg}^{0}\to 0
\end{eqnarray}
considered in the introduction, which will explain its relation to the string theory. 
In particular, this will give an embedding of \rf{cmplxforms} into the superspace with natural grading. 
In order to do this, we introduce odd variables $c_{\pm 1}$, $c_0$ and $D$ ($D$ is the dimension of the space-time) even variables $q^{\mu}$, $\mu=0,..., D-1$. Let us consider the space $\mF_{\mathfrak{g}}$, spanned by the elements of the following kind:
\begin{eqnarray}\label{frealiz1}
&&\rho_{u}=u(x), \quad \phi_{\mathbf{A}}=-ic_1A_{\mu}(x)q^{\mu}-c_0\p_{\mu}A^{\mu}(x),\nonumber\\ 
&&\psi_{\mathbf{W}}=-ic_1c_0 W_{\mu}(x)q^{\mu}, \quad \chi_{a}=2c_1c_0 c_{-1} a(x),
\end{eqnarray}  
which are associated with 0-forms $u$, $a$ and 1-forms $\mA$, $\mW$. The grading in this space 
is given by the number of odd $c$-variables (known as $ghost$ $number$) and can be realized by means of the operator $N_g=\sum^{n=1}_{n=-1}c_n\frac{\p}{\p c_n}$. It is easy to see that the space under consideration is invariant under the action of the following operator:
\begin{eqnarray}\label{diff}
Q=\sum^{1}_{n=-1}c_ns_{-n}-2c_{-1}c_1\frac{\p}{\p c_0},
\end{eqnarray}
where \footnote{This operator is the reduction of the usual BRST operator in the open string theory: $c_n$ correspond to the 
modes of c-ghost, the derivative with respect to $c_0$ is simply the $0$-th mode of $b$-ghost.}
\begin{eqnarray}
s_0=-2\frac{\p^2}{\p x^{\mu}\p x_{\mu}},\quad s_1=-i2\frac{\p^2}{\p x^{\mu}\p q_{\mu}},\quad 
s_{-1}=-i2q^{\mu}\frac{\p}{\p x^{\mu}}.
\end{eqnarray} 
Namely, the differential $Q$ acts on the elements of $\mF_{\mg}$ in the following way: 
\begin{eqnarray}\label{frealiz2}
&&0\to\mg\xrightarrow{i}\mathcal{F}_{\mg}^{0}\xrightarrow{Q}\mathcal{F}_{\mg}^{1}
\xrightarrow{Q}\mathcal{F}_{\mg}^{2}\xrightarrow{Q}\mathcal{F}_{\mg}^{3}\to 0,\nonumber\\
&&Q\rho_{u}=2\phi_{\ud u},\quad Q\phi_{\mathbf{A}}=2\psi_{\m \mA},\quad Q\psi_{\mathbf{W}}=-\chi_{\di \mW}, 
\quad Q\chi_{a}=0,
\end{eqnarray}
where the space $\mF_{\mg}^i$ $(i=0,1,2,3)$ consists elements of ghost number $i$. Therefore, we see that complex $(\mF_{\mg}^*,Q)$ gives a realization of the complex \rf{cmplxforms}. 
In the following we will refer to $\rf{cmplxforms}$ and $(\mF_{\mg}^*,Q)$ as the Yang-Mills chain complex. 

Next, we define the inner product on the chain complex $\mF_{\mg}$. 
Let $\Psi=\rho_{u}+\phi_{\mathbf{A}}+\psi_{\mathbf{U}}+\chi_{a}$ and 
$\Phi=\rho_{v}+\phi_{\mathbf{B}}+\psi_{\mathbf{V}}+\chi_{b}$. The inner product is given by the following formula:
\begin{eqnarray}
&&\langle\Psi,\Phi\rangle=\nonumber\\
&&\int d^Dx((\mA,\mathbf{V})(x)+(\mU,\mB)(x)-2(u(x),b(x))_K
-2(a(x),v(x))_K).
\end{eqnarray} 
This product has a very important property which shows how it behaves under the action of the differential $Q$. 

\vspace{3mm}

\noindent{\bf Proposition 2.1.} \cite{bvym} {\it Let $\Phi, \Psi\in \mF_{\mg}$ be of ghost numbers $n_{\Phi}$, 
$n_{\Psi}$. The following relation holds:
\begin{eqnarray}
\langle Q\Phi, \Psi\rangle=-(-1)^{n_{\Phi}n_{\Psi}}\langle Q\Psi, \Phi\rangle.
\end{eqnarray} }
{\bf 2.2. The Yang-Mills $L_{\infty}$ algebra.} First, we define the graded antisymmetric bilinear and trilinear algebraic operations which generate the homotopy Lie algebra.

\vspace{3mm}

\noindent{\bf Definition 2.1.} \cite{ym},\cite{bvym} {\it We define the graded (w.r.t. to the ghost number) 
antisymmetric bilinear and trilinear operations  
\begin{eqnarray}
&&[\cdot, \cdot]_h: \mathcal{F}^i_{\mathfrak{g}}\otimes \mathcal{F}^j_{\mathfrak{g}}\to \mathcal{F}^{i+j}_{\mathfrak{g}},\\
&&[\cdot, \cdot, \cdot]_h: \mathcal{F}^i_{\mathfrak{g}}\otimes \mathcal{F}^j_{\mathfrak{g}}\otimes \mathcal{F}^k_{\mathfrak{g}}\to 
\mathcal{F}^{i+j+k-1}_{\mathfrak{g}}
\end{eqnarray}
by the following relations on the elements of 
$\mathcal{F}_{\mathfrak{g}}$} : 
\begin{eqnarray}\label{ymha}
&&[\rho_{u},\rho_{v}]_h=2\rho_{[u,v]}, \quad
[\rho_{u},\phi_{\mathbf{A}}]_h=2\phi_{[u,\mathbf{A}]}, \quad 
[\rho_{u},\psi_{\mathbf{W}}]_h=2\phi_{[u,\mathbf{W}]}, \\
&&[\rho_u, \chi_a]_h=2\chi_{[u,a]},\quad[\phi_{\mathbf{A}},\phi_{\mathbf{B}}]_h=2\phi_{\{\mathbf{A},\mB\}}, \quad [\phi_{\mathbf{A}},\psi_{\mathbf{W}}]_h= 
-\chi_{\mathbf{A}\cdot\mW},\nonumber\\
&&[\phi_{\mathbf{A}},\phi_{\mathbf{B}} ,\phi_{\mathbf{C}} ]_h=2\psi_{\{\mA,\mB,\mC\}}\nonumber,
\end{eqnarray}
{\it where $\rho_u$, $\rho_v \in \mF^0_{\mathfrak{g}}$, $\phi_{\mathbf{A}}$, $\phi_{\mathbf{B}}$, 
$\phi_{\mathbf{C}}\in \mF^1_{\mathfrak{g}}$, 
$\psi_{\mathbf{W}}\in \mF^2_{\mathfrak{g}}$, $\chi_a\in \mF^3_{\mathfrak{g}}$ and the trilinear operation is nonzero only in the case when all arguments belong to $\mF_{\mg}^1$.}

\vspace{5mm}

\noindent{\bf Remark.} One can see from the Definition 2.1. that in the most of cases bilinear operation reduces to the commutator of the corresponding $\mathfrak{g}$-valued differential forms (we see that for example $\rho$ can be interpreted as a Lie morphism (up to a factor of 2) and $\phi$ is map of Lie modules.). The only "nontrivial" case, which prevents $[\cdot,\cdot]_h$ to satisfy the Jacobi identity (see below) is the one restricted to $\mF^1_{\mathfrak{g}}$. 

\vspace{5mm}

In the next proposition, we explicitly write down the relations of the homotopy Lie algebra corresponding to these operations.

\vspace{3mm}

\noindent{\bf Proposition 2.2.} \cite{ym} {\it Let $a_1,a_2, a_3, b, c$ $\in$ $\mF_{\mathfrak{g}}$ be of ghost numbers 
$n_{a_1}$, $n_{a_2}$, $n_{a_3}$, $n_b$, $n_c$ respectively. The following relations hold:}
\begin{eqnarray}\label{rel}
&&Q[a_1,a_2]_h=[Q a_1,a_2]_h+(-1)^{n_{a_1}}[a_1,Q a_2]_h,\nonumber\\
&&Q[a_1,a_2, a_3]_h+[Q a_1,a_2, a_3]_h+(-1)^{n_{a_1}}[a_1,Q a_2, a_3]_h+\nonumber\\
&&(-1)^{n_{a_1}+n_{a_2}}[ a_1, a_2, Q a_3]_h+[a_1,[a_2, a_3]_h]_h-[[a_1,a_2]_h, a_3]_h-\nonumber\\
&&(-1)^{n_{a_1}n_{a_2}}[a_2,[a_1, a_3]_h]_h=0,\nonumber\\
&&[b,[a_1,a_2, a_3]_h]_h-(-1)^{n_b(n_{a_1}+n_{a_2}+n_{a_3})}[a_1,[a_2, a_3, b]_h]_h+\nonumber\\
&&(-1)^{n_{a_2}(n_{b}+n_{a_1})}[a_2,[b,a_1, a_3]_h]_h-(-1)^{n_{a_3}(n_{a_1}+n_{a_2}+n_{b})}
[a_3,[b, a_1,a_2]_h]_h\nonumber\\
&&=[[b,a_1]_h,a_2, a_3]_h+(-1)^{n_{a_1}n_{b}}[a_1,[b,a_2]_h, a_3]_h+\nonumber\\
&&(-1)^{(n_{a_1}+n_{a_2})n_{b}}[a_1,a_2, [b,a_3]_h]_h.
\end{eqnarray}

\noindent{\bf Remark.} The relations in  the Proposition 2.2. can be described in the following way. The first one expresses the fact that $Q$ is a derivation for the bilinear operation $[\cdot ,\cdot ]_h$. The second relation is the homotopy Jacobi identity  for $[\cdot ,\cdot ]_h$, i.e.  $[\cdot ,\cdot ]_h$ satisfies the Jacobi identity up to homotopy given by the trilinear operation $[\cdot ,\cdot ,\cdot]_h$. 
The third relation gives the "higher" Jacobi  identity between bilinear and trilinear operations. 
 
 \vspace{3mm}
 
\noindent In order to rewrite the Yang-Mills action in the Homotopy Chern-Simons form, we need to define the following multilinear forms (see e.g. \cite{mss},\cite{zwiebach} and references therein).

\vspace{3mm}

\noindent{\bf Definition 2.2.}  {\it For any $a_1, a_2, a_3, a_4\in \mF_{\mg}$, one can define the following n-linear 
forms (n=2,3,4)
\begin{equation}
\{\cdot,..., \cdot\}_h: \mF_{\mg}\otimes ...\otimes \mF_{\mg}\to \mathbb{C} 
\end{equation}
in the following way:
\begin{eqnarray}
&&\{a_1,a_2\}_h=\langle Qa_1, a_2\rangle, \quad \{a_1,a_2,a_3\}_h=\langle [a_1, a_2]_h,a_3\rangle,\nonumber\\
&&\{a_1,a_2,a_3, a_4\}_h=\langle [a_1, a_2,a_3]_h,a_4\rangle.
\end{eqnarray}}
These bilinear operations satisfy the following property (see e.g. \cite{mss},\cite{zwiebach}).

\vspace{3mm}

\noindent {\bf Proposition 2.3.} {\it The multilinear products, introduced in Definition 2.2., are graded 
antisymmetric, i.e.
\begin{eqnarray}
\{a_1,...,a_i,a_{i+1},..., a_n\}_h=-(-1)^{n_{a_i}n_{a_{i+1}}}\{a_1,...,a_{i+1},a_{i},..., a_n\}_h.
\end{eqnarray}}

\noindent {\bf 2.3. The (BV) Yang-Mills as a Homotopy Chern-Simons.} Now we are ready to formulate the physical applications of 
the formalism we considered in the first two subsections. Namely, first we rewrite Yang-Mills equations as generalized 
Maurer-Cartan equations for the $L_{\infty}$ algebra, considered in the previous subsection. 

\vspace{3mm}

\noindent{\bf Proposition 2.4.} \cite{ym},\cite{bvym} {\it Let $\phi_{\mA}$ be the element of $\mF^1_{\mathfrak{g}}$ 
associated with 1-form $\mA=A_{\mu}dx^{\mu}$ and 
$\rho_u$ be the element of $\mF^0_{\mathfrak{g}}$ associated with a Lie algebra-valued function $u(x)$. Then, the Yang-Mills equations for $\mA$ and its infinitesimal gauge transformations
\begin{eqnarray}\label{ym}
&&*\ud*{\bf F}+[\mA,{\bf F}]=0,  \quad {\bf F}=\ud\mA+\mA\wedge\mA\nonumber\\
&&\label{gt}\mA\to \mA+\epsilon(\ud u+[\mA,u])
\end{eqnarray}
can be rewritten as follows:
\begin{eqnarray}\label{mc}
&&Q\phi_{\mA}+\frac{1}{2!}[\phi_{\mA},\phi_{\mA}]_h+\frac{1}{3!}[\phi_{\mA},\phi_{\mA},\phi_{\mA}]_h=0,\\
&&\phi_{\mA}\to \phi_{\mA}+\frac{\epsilon}{2}(Q\rho_u +[\phi_{\mA},\rho_u]_h).
\end{eqnarray}
}
Therefore, the Yang-Mills action can be reformulated in the Homotopy Chern-Simons form.

\vspace{3mm}

\noindent{\bf Proposition 2.5.} {\it The Yang-Mills action 
\begin{eqnarray}
S_{YM}=1/2\int d^Dx(F_{\mu\nu}(x),F^{\mu\nu}(x))_K 
\end{eqnarray}
can be written as follows:
\begin{eqnarray}\label{cs}
&&S_{YM}=-\sum^4_{n=2}\frac{1}{n!}\{\phi_{\mA}^n\}_h=\nonumber\\
&&-\frac{1}{2}\langle Q\phi_{\mA},\phi_{\mA}\rangle-\frac{1}{6}\{\phi_{\mA},\phi_{\mA},\phi_{\mA}\}_h-\frac{1}{24}
\{\phi_{\mA},\phi_{\mA},\phi_{\mA}, \phi_{\mA}\}_h.
\end{eqnarray}
}
Our next task is to represent the Batalin-Vilkovisky version of the Yang-Mills action in the same way. 

In order to introduce ghosts, antifields, i.e. the fermion degrees of freedom, we 
consider the tensor product of our chain complex $(\mF_{\mg}^{\cdot},Q)$ with some Grassmann algebra $A$:  $A=\oplus_{i\in \mathbb{Z}}A^i$. If $\lambda^i\in A^i$ and 
$\lambda^j\in A^j$, then  $\lambda^i\lambda^j=(-1)^{ij}\lambda^j\lambda^i$.
 
Moreover, we introduce the following notation: if $\lambda\in A^i$, we will say that $\lambda$ is of 
{\it target space ghost number} $i$.
Therefore, it is reasonable to introduce the gradation w.r.t. the {\it total ghost number}, which 
is equal to the sum of the {\it worldsheet ghost number}, generated by the operator $N_g$ and this target space ghost 
number on the space $\mathcal{H}_{\mg}=\mF_{\mg}\otimes A$. Hence,
 if the element $\Phi\in \mathcal{H}^n_{\mg}$ (i.e. of total ghost number $n$), 
which is written in the form $\sum_s\Phi_s\otimes\xi_s$ such that $\Phi_s\in \mF_{\mg}$ of ghost number $n^w_s$ and $\xi_s\in A$ 
of ghost number $n^t_s$, 
then $n^w_s+n^t_s=n$ for all $s$. In the following, to simplify the notation, 
we will refer to the total ghost number simply as the ghost number.  
This way, one can consider a new infinite chain complex $(\mathcal{H}_{\mg}^{\cdot},Q)$:
\begin{eqnarray}\label{infcomplex}
...\xrightarrow{Q}\mathcal{H}_{\mg}^{-1}\xrightarrow{Q}\mathcal{H}_{\mg}^{0}\xrightarrow{Q}\mathcal{H}_{\mg}^{1}
\xrightarrow{Q}\mathcal{H}_{\mg}^{2}\xrightarrow{Q}...,
\end{eqnarray}
where $Q$ denotes the action of $Q\otimes 1$. The space $\mathcal{H}_{\mg}=\oplus_{i\in\mathbb{Z}}\mathcal{H}_{\mg}^{i}$ of this complex is therefore generated by the elements of the 
form $\rho_u$, $\phi_{\mA}$, $\psi_{\mW}$, 
$\chi_a$, which are associated with the functions and 1-forms, which take values in $\mg \otimes A$. 
We can extend the algebraic structures, defined for the complex $\mF_{\mg}$ in section 2, to the 
space of the complex $\mathcal{H}_{\mg}$. 
For example, one can extend the operation $\langle \cdot, \cdot \rangle$ to be graded 
symmetric on the space $\mH_{\mg}$ 
with respect to the total ghost number. However, then it will take values in $A$ \cite{bvym}. 

Now we show that the BV Yang-Mills action can be rewritten as a Homotopy Chern-Simons action.

\vspace{3mm}

\noindent
{\bf Proposition 2.6.} {\it Consider $\Phi=\Phi(\omega, \omega^*, \mA, \mA^*)\in \mH^1_{\mg}$ 
such that
\begin{eqnarray}\label{field}
\Phi(\omega, \omega^*, \mA, \mA^*)=\rho_{\omega}+\phi_{\mathbf{A}}-\psi_{\mathbf{A^*}}-1/2\chi_{\omega^*}.
\end{eqnarray}
Then, the Homotopy Chern-Simons (HCS) action
\begin{eqnarray}
&&S_{HCS}=-\sum^4_{n=2}\frac{1}{n!}\{\Phi^n\}_h=\nonumber\\
&&-\frac{1}{2}\langle Q\Phi,\Phi\rangle-\frac{1}{6}\{\Phi,\Phi,\Phi\}_h-\frac{1}{24}
\{\Phi,\Phi,\Phi, \Phi\}_h
\end{eqnarray}
coincides with the BV Yang-Mills action
\begin{eqnarray}
&&S_{BVYM}=S_{YM}[\mA]+\\
&&2\int d^D x(\p_{\mu}\omega(x)+[A_{\mu}(x),\omega(x)],A^{*\mu}(x))_K-([\omega(x),\omega(x)],\omega^*(x))_K).\nonumber
\end{eqnarray}
} 

\noindent{\bf Remark.} The field $\omega$ is usually called $the$ $ghost$ field and the fields 
$\mA^*$, $\omega^*$ are called $antifields$ of the gauge field and the ghost field correspondingly. 

\section{Antighost and Gauge Fixing}
{\bf 3.1. The Yang-Mills complex with antighost field.}  
Let us consider the elements $\phi_a$ and $\psi_b$, related to 
$\mg$-valued functions $a(x)$ and $b(x)$: 
\begin{eqnarray}
\phi_a=c_0a(x), \quad \psi_b=2c_1c_{-1}b(x)-2ic_1c_0\p_{\mu}b(x)q^{\mu}
\end{eqnarray} 
such that 
\begin{eqnarray}\label{formext}
Q\phi_a=\psi_{a},\quad  Q\psi_b=0.
\end{eqnarray} 
Let us define the spaces $\mathcal{G}_{\mg}^{1}$ elements of type $\phi_a$ and $\mathcal{G}_{\mg}^{2}$ elements 
of type $\psi_b$ (the superscripts are given in accordance to the values of the grading operator). 
In this way, we get the complex $(\mG_{\mg}^{\cdot},Q)$:
\begin{eqnarray}\label{agcomplex}
0\to\mathcal{G}_{\mg}^{1}\xrightarrow{Q}\mathcal{G}_{\mg}^{2}\to 0.
\end{eqnarray}
Therefore, one can define a direct sum of the appropriate complex with the Yang-Mills one: 
\begin{eqnarray}\label{agymcomplex}
0\to\mg\xrightarrow{i}\mathcal{F}_{ag}^{0}\xrightarrow{Q}\mathcal{F}_{ag}^{1}
\xrightarrow{Q}\mathcal{F}_{ag}^{2}\xrightarrow{Q}\mathcal{F}_{ag}^{3}\to 0
\end{eqnarray}
such that 
\begin{eqnarray}
&&\mathcal{F}_{ag}^{i}=\mathcal{F}_{\mg}^{i}\oplus \mathcal{G}_{\mg}^{i} \quad (i=1,2),\nonumber\\  
&&\mathcal{F}_{ag}^{j}=\mathcal{F}_{\mg}^{j} \quad (j=0,3). 
\end{eqnarray}
In other words,  \rf{formext} together with \rf{frealiz1}, \rf{frealiz2} gives a realization of the following extension of the Yang-Mills complex:
\[
\xymatrixcolsep{20pt}
\xymatrixrowsep{0pt}
\xymatrix{
0\ar[r]&\Omega^0_{\mathfrak{g}} \ar[r]^{\ud} & \Omega^1_{\mathfrak{g}} \ar[r]^{\m}  &\Omega^1_{\mathfrak{g}} \ar[r]^{\di} & \Omega^0_{\mathfrak{g}}\ar[r]&0\\
 && \bigoplus & \bigoplus     &&\\
 &0\ar[r]& \Omega^0_{\mathfrak{g}}\ar[r]^{id}   & \Omega^0_{\mathfrak{g}}\ar[r]&0
}
\]
We will refer to this complex as the $Yang$-$Mills$ $complex$ $with$ $antighost$ (below we will 
show that the elements of the lower subcomplex are related to what is known as antighost and its 
antifield) and denote it as $(\mathcal{F}^{\cdot}_{ag},Q)$. Using the analogy with BPZ inner product, one can define the nondegenerate inner product on this complex such that 
it is invariant under the action of the operator $Q$, as in Proposition 2.1. Namely, for two elements 
$\Phi=\rho_{u}+\phi_{\mathbf{A}}+\phi_a+ \psi_{\mathbf{W}}+\psi_b+\chi_{v}$ and 
$\Phi'=\rho_{u'}+\phi_{\mathbf{A}'}+\phi_{a'}+ \psi_{\mathbf{W}'}+\psi_{b'}+\chi_{v'}$, the inner product 
is given by:
\begin{eqnarray}\label{ipag}
&&\langle\Phi,\Phi'\rangle=\int d^Dx((\mA,\mathbf{W}')(x)+(\mW,\mA')(x)-\\
&&2(u(x),v'(x))_K-2(v(x),u'(x))_K-2(a(x),b'(x))_K-2(b(x),a'(x))_K.\nonumber
\end{eqnarray} 
One can easily generalize multilinear operations $[\cdot,...,\cdot]_h$
in such a way that they are zero when one of the 
arguments belongs to $\mG^i_{\mg}$. Similarly one can generalize the definition of graded antisymmetric 
multilinear forms $\{\cdot,...,\cdot\}_h$ by means of the inner product \rf{ipag}. 
In the next subsection, we will use this extension of Yang-Mills homotopy algebra in order to define 
the gauge fixed Yang-Mills theory. 

\vspace{3mm}

\noindent{\bf 3.2. The gauge fixing operator and BV Yang-Mills.} 
One can define a nilpotent operator which acts ``backwards'' w.r.t. the action of $Q$ and has zero cohomology. 
Namely, this operator is $\mathbf{b}=\frac{1}{2}\frac{\p}{\p c_0}$. Its action on the elements of $\mF_{ag}$ is 
as follows:
\begin{eqnarray}
&&\mb\rho_u=0, \quad \mb\phi_{\mA}=-\frac{1}{2}\rho_{\di \mA}, \quad \mb\phi_{a}=\frac{1}{2}\rho_{a},\\ 
&&\mb\psi_{\mW}=\frac{1}{2}(\phi_{\mW}+\phi_{\di \mW}), \quad \mb\psi_b=-\phi_{\d b}-\phi_{\p^{\mu}\p_{\mu}b}, 
\quad \mb \chi_v=-\frac{1}{2}\psi_v+\psi_{\d v}.\nonumber
\end{eqnarray}
Moreover, the following relation holds:
\begin{eqnarray}
[Q,\mb]+\Delta=0,
\end{eqnarray}
where $\Delta=\p_{\mu}\p^{\mu}$ is a Laplacian. Therefore $\mathbf{b}$ is a contracting homotopy 
for $\Delta$ and we obtain the following proposition (cf. the standard Hodge decomposition for differential forms).\\

\noindent{\bf Proposition 3.1.} {\it The space of complex $(\mF^{\cdot}_{ag},Q)$ decomposes into the direct sum:
\begin{eqnarray}\label{decomp}
\mF_{ag}=\oplus^3_{i=0}\mF^i_{ag}=Im \mb\oplus Im Q\oplus Ker \Delta.
\end{eqnarray}}

As in the case of the Yang-Mills complex, one can consider the complex $(\mH^{\cdot}_{ag},Q)$, where 
$\mH^{\cdot}_{ag}=\mF^{\cdot}_{ag}\otimes A$, $Q$ acts as $Q\otimes 1$ and $A$ is a 
Grassmann algebra with grading, which corresponds to the target space ghost number. The Maurer-Cartan element in $\mH^{\cdot}_{ag}$ (the element of total ghost number 1) is: 
\begin{eqnarray}
\Phi_{ag}=\rho_{\omega}+\phi_{\b \omega^*}+\phi_{\mA}-1/2\psi_{\b\omega}-
\psi_{\mA^*}-1/2\chi_{\omega^*}, 
\end{eqnarray}
where $\b \omega(x)$ is of ghost number equal to $-1$ and is called the $antighost$ field \cite{weinberg}.  
Here $\b \omega^*(x)$ is its antifield of ghost number zero, which is sometimes called the Nakanishi-Lautrup field. 
It is easy to check that the following proposition holds.\\

\noindent{\bf Proposition 3.2.} {\it The condition $\mb\Phi_{ag}=0$ gives the following relations between fields and antifields:
\begin{eqnarray}
\b \omega^*=\di\mA, \quad \mA^*=\d \b \omega, \quad \omega^*=0.
\end{eqnarray}}

\noindent{\bf Remark.} (i) The condition $\mb\Phi_{ag}=0$ is equivalent 
to the condition $\Phi_{ag}\in Im \mb$ due to the fact 
that this operator has zero cohomology when acting on the space of the complex $(\mH^{\cdot}_{ag},Q)$. (ii) This condition in terms of fields and antifields corresponds to one of the most popular choices of the Lagrangian submanifold w.r.t. the Yang-Mills BV structure, namely the Lorenz gauge condition. 

\vspace{3mm}

\noindent If we plug $\Phi_{ag}$ into the homotopy Chern-Simons action:
\begin{eqnarray}\label{hcsag}
&&S_{HCS}[\Phi_{ag}]=-\frac{1}{2}\langle\Phi_{ag}, Q\Phi_{ag}\rangle-\frac{1}{3!}\{\Phi_{ag},\Phi_{ag},\Phi_{ag}\}_h-\nonumber\\
&&\frac{1}{4!}\{\Phi_{ag},\Phi_{ag},\Phi_{ag},\Phi_{ag}\}_h,
\end{eqnarray}
we find that it is equivalent to 
\begin{eqnarray}
S_{HCS}[\Phi_{ag}]=S_{BVYM}(\omega,\mA,\mA^*,\omega^*)+(\b \omega^*)^2.
\end{eqnarray}
Therefore, with the condition $\mb\Phi_{ag}=0$, \rf{hcsag} looks as follows:
\begin{eqnarray}
&&S^{gf}_{BVYM}=\\
&&\int d^Dx(\frac{1}{2}(F_{\mu\nu}(x)F^{\mu\nu}(x))_K+2(D_{\mu}\omega(x),\p_{\mu}\b \omega(x))_K+(\p_{\mu}A^{\mu},\p_{\nu}A^{\nu})_K),\nonumber
\end{eqnarray}
which coincides with the Yang-Mills gauge fixed action in the Lorenz gauge. Thus the generating functional for the quantum theory 
should be written in such notation as:
\begin{eqnarray}\label{genfun}
Z[J]=\int_{\mb \Phi_{ag}=0}[d\Phi_{ag}]e^{-S_{HCS}[\Phi_{ag}]+\langle\Phi_{ag},J\rangle}.
\end{eqnarray}

\noindent{\bf 3.3. A nontrivial generalization of the Yang-Mills $L_{\infty}$ algebra by means of 
antighost fields.} 
In the previous subsection, we used the simplest possible generalization of the Yang-Mills homotopy Lie algebra for the complex $(\mF_{ag}^{\cdot},Q)$. 
Namely, we considered multilinear operations to be equal to 0 when one of the arguments belongs to 
the complex $(\mG_{\mg}^{\cdot},Q)$.

However, the construction of the bilinear operations in the case of the usual Yang-Mills chain complex 
was based on the algebraic operation constructed from OPEs of certain operators in boundary CFT of 
open string theory \cite{zeit3}. In our case, 
let us consider the operators 
\begin{eqnarray}
\boldsymbol{{\phi}}^{(0)}_a=\p c a(X), \quad \boldsymbol{{\phi}}^{(0)}_b=c\p^2 c b(X),
\end{eqnarray}
where $c$ is the $c$-ghost from the string $b$-$c$ system \cite{pol} and $X$ is the usual string coordinate. These operators can be identified  with the elements  $\phi_a=c_0a(x)$ and 
$\psi'_b=\psi_b-2\psi_{\d b}=2c_1c_{-1}b(x)$ from the complex $(\mG^{\cdot}_{\mg},Q)$. 
Let us neglect the explicit dependence of $\alpha'$ in the operator product expansion, namely we put 
$\alpha'=2$. Then, considering the bilinear operation of \cite{zeit3} acting on operators, corresponding to states from $(\mF^{\cdot}_{ag},Q)$, one can see that the contribution of elements from $(\mG^{\cdot}_{\mg},Q)$  becomes nontrivial. Then, as before, one can construct a trilinear operation on $(\mF^{\cdot}_{ag},Q)$ in order to make the relations of $L_{\infty}$ algebra hold. The result can be summarised as follows. 

\vspace{3mm}

\noindent {\bf Definition 3.1.} (i) {\it We define a graded (w.r.t. to the ghost number) antisymmetric
bilinear operation
\begin{eqnarray}
[\cdot, \cdot]_h: \mathcal{F}_{ag}^i\otimes \mathcal{F}_{ag}^j\to \mathcal{F}_{ag}^{i+j},
\end{eqnarray}
by the following relations (accompanied with those of \rf{ymha}) on the elements of 
$\mathcal{F}_{ag}$}: 
\begin{eqnarray}\label{relag1}
&&[\rho_u,\phi_{a}]_h=2\phi_{[u,a]},\quad [\rho_u,\psi'_{a}]_h=2\psi'_{[u,a]},\quad [\phi_{\mA},\phi_{a}]_h=2\psi_{[\mA,a]},
\nonumber\\
&& [\phi_{\mA},\psi'_{a}]_h=2\chi_{*\ud[*\mA,a]+*[\mA,*\ud a]}, \quad 
[\psi_{\mW},\phi_{a}]_h=0, \quad [\phi_a,\phi_b]_h=0.
\end{eqnarray} 
(ii) {\it The graded (w.r.t. to the ghost number) trilinear operation on $\mF_{ag}$ 
\begin{eqnarray}
[\cdot, \cdot, \cdot]_h: \mathcal{F}_{ag}^i\otimes \mathcal{F}_{ag}^j\otimes \mathcal{F}_{ag}^k \to \mathcal{F}_{ag}^{i+j+k-1}
\end{eqnarray}
is defined to be nonzero either if all three elements belong to $\mathcal{F}^1_{\mg}$ as it was in \rf{ymha} or 
if two of the arguments belong to $\mathcal{F}^1_{\mg}$ and the third belongs to $\mG^2_{\mg}$}:
\begin{eqnarray}\label{relag2}
[\phi_{\mA},\phi_{\mB}, \psi'_a]_h=2\chi_{*[\mA,*[\mB,a]]+*[\mB,*[\mA,a]]}.
\end{eqnarray}
The following proposition shows that these operations generate the $L_{\infty}$ algebra on $\mF_{ag}$.

\vspace{3mm} 

\noindent {\bf Proposition 3.3.} {\it The extension of bilinear and trilinear operations given by the relations 
\rf{relag1}, \rf{relag2} provides that these operations satisfy the homotopy Lie algebra relations \rf{rel} on the elements of  the complex $(\mF_{ag}^{\cdot}, Q)$.}\\
{\bf Proof.} 
First, we will prove the relation 
\begin{eqnarray}\label{qb}
Q[a_1,a_2]_h=[Q a_1,a_2]_h+(-1)^{n_{a_1}}[a_1,Q a_2]_h.
\end{eqnarray} 
We have already proved it for $a_1, a_2\in \mF_{\mathfrak{g}}$, so we need to check only the cases when 
at least one of them belongs to $\mG_{\mathfrak{g}}$. So, let $a_1=\rho_u$, $a_2=\phi_{a}$. Then 
\begin{eqnarray}
&&[Q\rho_u,\phi_a]_h+[\rho_u,Q\phi_a]_h=\nonumber\\
&&[2\phi_{\ud u} ,\phi_a]_h+[\rho_u,\psi'_a+2\psi_{\ud a}]_h=4\psi_{\ud [u,a]}+2\psi'_{[u,a]}=2Q\phi_{[u,a]}=\nonumber\\
&&Q[\rho_u,\phi_a]_h.
\end{eqnarray}
For $a_1=\rho_u$, $a_2=\psi_{a}$, we have:
\begin{eqnarray}
&&[Q\rho_u,\psi'_a]_h+[\rho_u,Q\psi'_a]_h=\nonumber\\
&&2[\phi_{\ud u}, \psi_a]_h+[\rho_u,2\chi_{\p_{\mu}\p^{\mu}a}]_h=4\chi_{\p^{\mu}[\p_{\mu}u,a]+
[\p_{\mu}u,\p^{\mu}a]+[u,\p_{\mu}\p^{\mu}a]}= \nonumber\\
&&2Q[\rho_u,\psi'_a]_h.
\end{eqnarray}
Let $a_1=\phi_\mA$, $a_2=\phi_{a}$. Then 
\begin{eqnarray}
&&Q[\phi_{\mA},\phi_a]_h=2Q\psi_{[\mA, a]}=-2\chi_{\di [\mA, a]},\nonumber\\
&&[Q\phi_{\mA}, \phi_a]_h-[\phi_{\mA}, Q\phi_a]_h=[2\psi_{\m \mA}, \phi_a]_h-[\phi_{\m \mA}, \psi_a+2\psi_{\ud a}]_h=\nonumber\\
&&-2\chi_{\p^{\mu}[A_{\mu}, a]}=-2\chi_{\di [\mA, a]}.
\end{eqnarray}
For all other possible values of $a_1, a_2$, the relation \rf{qb} is trivial.
Another relation we need to check is:
\begin{eqnarray}\label{jac}
&&Q[a_1,a_2, a_3]_h+[Q a_1,a_2, a_3]_h+(-1)^{n_{a_1}}[a_1,Q a_2, a_3]_h+\nonumber\\
&&(-1)^{n_{a_1}+n_{a_2}}[ a_1, a_2, Q a_3]_h+[a_1,[a_2, a_3]_h]_h-[[a_1,a_2]_h, a_3]_h-\nonumber\\
&&(-1)^{n_{a_1}n_{a_2}}[a_2,[a_1, a_3]_h]_h=0.
\end{eqnarray}
First, we consider the case when $a_1=\phi_{\mA}$, $a_2=\phi_{\mB}$, $a_3=\phi_a$. 
We see that 
\begin{eqnarray}
&&[\phi_{\mA},[\phi_{\mB}, \phi_a]_h]_h-[[\phi_{\mA},\phi_{\mB}]_h, \phi_a]_h+[\phi_{\mB}, [\phi_{\mA}, \phi_a]_h]_h=\nonumber\\
&&-2\chi_{[A_{\mu},[B^{\mu}, a]]+[B_{\mu},[A^{\mu}, a]]}.
\end{eqnarray}
At the same time 
\begin{eqnarray}
[\phi_{\mA},\phi_{\mB}, Q\phi_a]_h
=[\phi_{\mA},\phi_{\mB}, \psi'_a]_h=2\chi_{[A_{\mu},[B^{\mu}, a]]+[B_{\mu},[A^{\mu}, a]]}.
\end{eqnarray}
Therefore, the relation \rf{jac} holds in this case. 
The only nontrivial possibility for the values of $a_i$ is as follows: $a_1=\rho_u$, $a_2=\phi_{\mA}$, $a_3=\phi_a$. 
In this case, let us write down each term separately:
 \begin{eqnarray}
&&[\rho_u,[\phi_{\mA}, \psi'_a]_h]_h=4\chi_{[u, \p^{\mu}[A_{\mu},a]+[A^{\mu},\p_{\mu}a]]},\nonumber\\
&&[[\rho_u,\phi_{\mA}]_h, \psi'_a]_h=4\chi_{\p^{\mu}[u, [A_{\mu},a]]+[[u,A^{\mu}],\p_{\mu}a]},\nonumber\\
&&[\phi_{\mA},[\phi_u, \psi'_a]_h]_h=4\chi_{\p^{\mu}[A_{\mu}[u,a]]+[A^{\mu},\p_{\mu}[u,a]]}.
\end{eqnarray}
Now, taking the sum of these terms with the appropriate signs, we find:
\begin{eqnarray}
&&[\rho_u,[\phi_{\mA}, \psi'_a]_h]_h-[[\rho_u,\phi_{\mA}]_h, \psi'_a]_h
-[\phi_{\mA},[\phi_u, \psi'_a]_h]_h=\nonumber\\
&&-4\chi_{[A^{\mu},[\p^{\mu}u,a]]+[\p^{\mu}u,[A^{\mu}, a]]}=
-[Q\rho_u,\phi_{\mA}, \psi'_a]_h.
\end{eqnarray}
Thus, in this case, the relation \rf{jac} is also proven. For all other values of $a_1, a_2, a_3$ we did 
not check, the relation \rf{jac} reduces to the Jacobi identity and it is easy to see that it is satisfied.
The same applies to 
\begin{eqnarray}
&&[b,[a_1,a_2, a_3]_h]_h=[[b,a_1]_h,a_2, a_3]_h+(-1)^{n_{a_1}n_{b}}[a_1,[b,a_2]_h, a_3]_h+\nonumber\\
&&(-1)^{(n_{a_1}+n_{a_2})n_{b}}[a_1,a_2, [b,a_3]_h]_h,\nonumber\\
&&[[a_1,a_2, a_3]_h,b,c]_h=0.
\end{eqnarray}
Thus the proposition is proven. $\blacksquare$

\vspace{3mm}

\noindent However, this extension has one huge disadvantage. If one builds the multilinear forms $\{\cdot,..., \cdot\}_h$ by means of the 
inner product \rf{ipag}, one finds that these forms are $no$ $longer$ graded antisymmetric. Therefore,  one cannot build the 
homotopy Chern-Simons action (since in order to vary the action successfully, the multilinear forms need to be graded antisymmetric). At the same time, we obtained this extension putting $\alpha'=2$. Therefore, 
it might be incomplete and one should seek for further extensions which will lead to graded  antisymmetric forms $\{\cdot,..., \cdot\}_h$. We considered this particular extension just to give an example of what can happen at the higher orders in $\alpha'$.

\section{The Scalar Field}
In this section, we consider the scalar field with values in some reductive Lie algebra, coupled to the 
Yang- Mills field. 
We will show that the action for the resulting theory has also the homotopy Chern-Simons form and the equations 
of motion are equivalent to the generalized Maurer-Cartan equations in some extension of the Yang-Mills $L_{\infty}$ algebra.

\vspace{3mm}

\noindent{\bf 4.1. The Scalar Field extension of the Yang-Mills complex.} First of all, we will define an extension of the complex $(\mF^{\cdot}_{\mg},Q)$, extending it by the elements corresponding to a scalar field, which is similar to the one considered in section 3:
\[
\xymatrixcolsep{20pt}
\xymatrixrowsep{0pt}
\xymatrix{
0\ar[r]&\Omega^0_{\mathfrak{g}} \ar[r]^{\ud} & \Omega^1_{\mathfrak{g}} \ar[r]^{\m}  &\Omega^1_{\mathfrak{g}} \ar[r]^{*\ud*} & \Omega^0_{\mathfrak{g}}\ar[r]&0\\
 && \bigoplus & \bigoplus     &&\\
 &0\ar[r]& \Omega^0_{\mathfrak{g}}\ar[r]^{K}   & \Omega^0_{\mathfrak{g}}\ar[r]&0,
}
\]
where $K=\Delta-m^2$ is a Klein-Gordon operator.

The explicit realization of this complex, extending the one given in subsection 2.1. can be constructed as follows. Let 
us  introduce an odd element $d_1$, an even element $d_2$ and a grading operator $\t N_g=d_1\frac{\p}{\p d_1}+2d_2\frac{\p}{\p d_2}.$
Considering $\phi_a=d_1 a(x)$ and $\psi_b=d_2 b(x)$, where $a(x)$, $b(x)$ are Lie algebra valued functions, one can define an operator $\t Q=2Kd_2\frac{\p}{\p d_1}$, such that 
\begin{eqnarray}
\t Q\phi_a=2\psi_{Ka},\quad \t Q\psi_b=0.
\end{eqnarray} 
Let us introduce the spaces $\mathcal{K}_{\mg}^{1}$ elements of type $\phi_a$ and $\mathcal{K}_{\mg}^{2}$ elements of type $\psi_b$ (the superscripts are given in accordance to the values of the grading operator). 
In this way, we get the complex $(\mK^{\cdot},\t Q)$: 
$
0\to\mathcal{K}_{\mg}^{1}\xrightarrow{\t Q}\mathcal{K}_{\mg}^{2}\to 0.
$
Therefore, one can define a direct sum of the appropriate complex with the 
Yang-Mills one. Namely, we consider 
$Q^{tot}=Q+\t Q$ in the complex 
\begin{eqnarray}\label{sfymcomplex}
0\to\mg\xrightarrow{i}\mathcal{F}_{sf}^{0}\xrightarrow{Q^{tot}}\mathcal{F}_{sf}^{1}
\xrightarrow{Q^{tot}}\mathcal{F}_{sf}^{2}\xrightarrow{Q^{tot}}\mathcal{F}_{sf}^{3}\to 0
\end{eqnarray}
such that 
\begin{eqnarray}\label{dsum}
&&\mathcal{F}_{sf}^{i}=\mathcal{F}_{\mg}^{i}\oplus \mathcal{K}_{\mg}^{i} \quad (i=1,2),\nonumber\\  
&&\mathcal{F}_{sf}^{j}=\mathcal{F}_{\mg}^{j} \quad (j=0,3). 
\end{eqnarray}
In this section, we will denote $Q^{tot}$ as $Q$, $N_g^{tot}=N_g+\t N_g$ as $N_g$, and the complex 
\rf{sfymcomplex} as $(\mathcal{F}_{sf}^{\cdot},Q)$. 
In order to define the homotopy Chern-Simons action on the space of the above complex, we will also need an inner product which is invariant under the action of the differential. For two elements 
$\Phi=\rho_{u}+\phi_{\mathbf{A}}+\phi_a+ \psi_{\mathbf{W}}+\psi_b+\chi_{v}$ and 
$\Phi'=\rho_{u'}+\phi_{\mathbf{A}'}+\phi_{a'}+ \psi_{\mathbf{W}'}+\psi_{b'}+\chi_{v'}$, the pairing is given by:
\begin{eqnarray}\label{sfip}
&&\langle\Phi,\Phi'\rangle=\int d^Dx((\mA,\mathbf{W}')(x)+(\mW,\mA')(x)-\nonumber\\
&&2(u(x),v'(x))_K-2(v(x),u'(x))_K-(a(x),b'(x))_K-(b(x),a'(x))_K,
\end{eqnarray} 
and we see that the direct sum in \rf{dsum} is orthogonal. It is easy to check that this product satisfies the properties from Proposition 2.1.

\vspace{3mm}

\noindent{\bf 4.2. The $L_{\infty}$ algebra on the complex $(\mF^{\cdot}_{sf},Q)$.} Now we will appropriately define the bilinear 
and trilinear operations on the space of the chain complex $(\mF^{\cdot}_{sf},Q)$.

\vspace{3mm}
 
\noindent {\bf Definition 4.1.} {\it (i) The bilinear operation
\begin{eqnarray}
[\cdot, \cdot]_h: \mathcal{F}^i_{sf}\otimes \mathcal{F}^j_{sf}\to \mathcal{F}^{i+j}_{sf}
\end{eqnarray}
is defined by means of \rf{ymha} and the following relations:
\begin{eqnarray}\label{bisf}
&&[\rho_v,\phi_u]_h=2\phi_{[v,u]}, \quad [\rho_v,\psi_u]_h=2\psi_{[v,u]},\quad [\phi_u,\phi_v]_h=2\psi_{[u,\ud v]+[v,\ud u]}\nonumber\\
&& [\phi_u,\psi_v]_h=\chi_{[u,v]},\quad [\phi_u,\phi_{\mA}]_h=2\psi_{*\ud*[\mA,u]+*[\mA,*\ud u]}, \nonumber\\ 
&& [\phi_u,\psi_{\mW}]_h=0,\quad [\phi_{\mW},\psi_{u}]_h=0.
\end{eqnarray}
(ii) The trilinear operation 
\begin{eqnarray}
\label{3lin}[\cdot, \cdot, \cdot]_h: \mathcal{F}^i_{sf}\otimes \mathcal{F}^j_{sf}\otimes \mathcal{F}^k_{sf}\to 
\mathcal{F}^{i+j+k-1}_{sf}
\end{eqnarray}
is defined to be nonzero when all arguments belong to $\mF^1_{sf}$ 
and in addition to \rf{ymha}, the following relations hold:}
\begin{eqnarray}\label{trisf}
[\phi_u,\phi_v,\phi_{\mA}]=2\psi_{[u,[\mA,v]]+[v,[\mA,u]]}, \quad [\phi_{\mA},\phi_{\mB},\phi_v]=2\psi_{*[\mA,*[\mB,v]]+
*[\mB,*[\mA,v]]}.
\end{eqnarray}
In such a way, the following proposition takes place.

\vspace{3mm}

\noindent {\bf Proposition 4.1.} {\it The extension of bilinear and trilinear operations given by the relations 
\rf{bisf}, \rf{trisf} provides that these operations satisfy the homotopy Lie algebra relations \rf{rel} on the elements of the complex $(\mF_{sf}^{\cdot}, Q)$.}\\
{\bf Proof.} As in Proposition 3.3, we start proving from the relation   
\begin{eqnarray}\label{qbt}
Q[a_1,a_2]_h=[Q a_1,a_2]_h+(-1)^{n_{a_1}}[a_1,Q a_2]_h.
\end{eqnarray} 
Keeping in mind the fact that the relation \rf{rel} for elements from $(\mF_{\mathfrak{g}}^{\cdot}, \mQ)$, we observe that the nontrivial cases are those when $a_1=\rho_v$, $a_2=\phi_u$ and  $a_1=\phi_u$, $a_2=\phi_v$ for some $\mg$-valued functions $u$ and $v$. 
Consider the case when $a_1=\rho_v$, $a_2=\phi_u$:
\begin{eqnarray}
&&[Q\rho_v,\phi_u]_h+[\rho_v,Q\phi_u]_h=2[\phi_{\ud v}, \phi_u]_h+2[\rho_v,\psi_{\p_{\mu}\p^{\mu}u-m^2u}]_h=\nonumber\\
&&4\psi_{\p^{\mu}\p_{\mu}[v,u]-m^2[v,u]}=Q[\rho_v,\phi_u]_h. 
\end{eqnarray} 
One can easily show that this relation also holds when $a_1=\phi_u$, $a_2=\phi_v$:
\begin{eqnarray}
&&[Q\phi_u,\phi_v]_h-[\phi_u,Q\phi_v]_h=2[\psi_{\p_{\mu}\p^{\mu}u-m^2u},\phi_v]_h-2[\phi_u,\psi_{\p_{\mu}\p^{\mu}v-m^2v}]_h=\nonumber\\
&&2\chi_{[\p^{\mu}\p_{\mu}u,v]-[u,\p^{\mu}\p_{\mu}v]}=2Q\psi_{[u,\p_{\mu}v]+[v,\p_{\mu}u]}=Q[\phi_u,\phi_v]_h.
\end{eqnarray} 
For all other values of $a_1, a_2$, the relation \rf{qbt} is either already checked (when $a_1,a_2\in \mF_{\mathfrak{g}}$) or trivial. 
Now, let's switch to the most interesting relation in our algebra, the one which represents super-Jacobi 
identity that holds  up to homotopy operator, namely 
\begin{eqnarray}\label{jact}
&&Q[a_1,a_2, a_3]_h+[Q a_1,a_2, a_3]_h+(-1)^{n_{a_1}}[a_1,Q a_2, a_3]_h+\nonumber\\
&&(-1)^{n_{a_1}+n_{a_2}}[ a_1, a_2, Q a_3]_h+[a_1,[a_2, a_3]_h]_h-[[a_1,a_2]_h, a_3]_h-\nonumber\\
&&(-1)^{n_{a_1}n_{a_2}}[a_2,[a_1, a_3]_h]_h=0.
\end{eqnarray}
There are three nontrivial cases which are not yet verified: (i) 
$a_1=\rho_u$, $a_2=\phi_{\mA}$, $a_3=\phi_v$, (ii) $a_1=\phi_u$, 
$a_2=\phi_v$, $a_3=\rho_a$, (iii) $a_1=\phi_u$, $a_2=\phi_{v}$, $a_3=\phi_{\mA}$. We will prove the relation \rf{jact} 
for all these cases.\\ 
(i) 
Let's write down all the terms explicitly: 
\begin{eqnarray}\label{tt}
&&[\rho_u,[\phi_{\mA}, \phi_v]_h]_h=-4\psi_{2[u,[\p_{\mu}v,A^{\mu}]]+[u,[v,\p^{\mu}A_{\mu}]]},\nonumber\\
&&[[\rho_u,\phi_{\mA}]_h, \phi_v]_h=-4\psi_{2[\p_{\mu}v,[u,A^{\mu}]]+[v,\p^{\mu}[u,A_{\mu}]},\nonumber\\
&&[\phi_{\mA}, [\rho_u,\phi_v]_h]_h=-4\psi_{2[\p_{\mu}[u,v],A^{\mu}]+[[u,v],\p^{\mu}A_{\mu}]}.
\end{eqnarray}
Summing \rf{tt} with the appropriate coefficients, we find that
\begin{eqnarray}
&&[\rho_u,[\phi_{\mA}, \phi_v]_h]_h-[[\rho_u,\phi_{\mA}]_h, \phi_v]_h-[\phi_{\mA}, [\rho_u,\phi_v]_h]_h=\nonumber\\
&&-2\psi_{2[\p_{\mu}u, [A^{\mu}, v]]+2[\p_{\mu}u, [A^{\mu}, v]]}=-[\phi_{2\ud u},\phi_{\mA},\phi_v]_h=
-[Q\rho_u,\phi_{\mA},\phi_v]_h.
\end{eqnarray}
(ii) Here 
\begin{eqnarray}
&&[\phi_u,\phi_v,Q\rho_a]=2[\phi_u,\phi_v,\phi_{\d a}]=-4\psi_{[u,[v,\d a]]+[v,[u,\d a]]},\nonumber\\
&&[\phi_u,[\phi_v,\rho_a]]=2[\phi_u,\phi_{[v,a]}]=4\psi_{[u,\d[v,a]]+[[v,a],\d u]},\nonumber\\
&&[\phi_v,[\phi_u,\rho_a]]=4\psi_{[v,\d [u,a]]+[[u,a],\d v]},\nonumber\\
&&[[\phi_u,\phi_v],\rho_a]=4\psi_{[[u,\d v]+[v,\d u],a]}.
\end{eqnarray}
Summing all terms, we find the desired relation:
\begin{eqnarray}
[\phi_u,\phi_v,Q\rho_a]_h+[\phi_u,[\phi_v,\rho_a]_h]_h-[[\phi_u,\phi_v],\rho_a]_h+
[\phi_v,[\phi_u,\rho_a]_h]_h=0.
\end{eqnarray}
(iii) Similarly, we write down all terms in this case:
\begin{eqnarray}
&&[\phi_u,[\phi_v,\phi_{\mA}]_h]_h=2\chi_{[u,\p_{\mu}[A^{\mu},v]]+[u,[\p_{\mu}A^{\mu},v]]},\\
&&[\phi_v,[\phi_u,\phi_{\mA}]_h]_h=2\chi_{[v,\p_{\mu}[A^{\mu},u]]+[v,[\p_{\mu}A^{\mu},u]]},\nonumber\\
&&[[\phi_u,\phi_v]_h,\phi_{\mA}]_h=2\chi_{[A^{\mu},[u,\p_{\mu}v]]+[A^{\mu},[v,\p_{\mu}u]]},\nonumber\\
&&Q[\phi_u,\phi_v,\phi_{\mA}]_h=\nonumber\\
&&2\chi_{[\p^{\mu}u,[v,A_{\mu}]]+[\p^{\mu}v,[u,A_{\mu}]]+[\p^{\mu}v,[u,A_{\mu}]]+
[v,[\p^{\mu}u,A_{\mu}]]+[u,[v,\p^{\mu}A_{\mu}]]+[v,[u,\p^{\mu}A_{\mu}]]}\nonumber
\end{eqnarray}
and summing them, we find:
\begin{eqnarray}
Q[\phi_u,\phi_v,\phi_{\mA}]_h+[\phi_u,[\phi_v,\phi_{\mA}]_h]_h-[[\phi_u,\phi_v]_h,\phi_{\mA}]_h+
[\phi_v,[\phi_u,\phi_{\mA}]_h]_h=0.
\end{eqnarray}
Thus the relation \rf{jact} is proven. The other relations are either trivial or easily follow from Jacobi identity. Proposition is proved. $\blacksquare$

\vspace{3mm}

\noindent{\bf 4.3. The generalized Maurer-Cartan, a homotopy Chern-Simons and the scalar field coupled to the Yang-Mills field.} 
Now we are ready to relate the above construction of the homotopy Lie algebra to the theory of the coupled scalar and 
gauge fields.
 
\vspace{3mm}

\noindent{\bf Proposition 4.2.} {\it The generalized Maurer-Cartan equation for 
the element $\Phi_{\varphi,\mA}=\phi_{\mA}+\phi_{i\varphi}$ from the complex 
$(\mF^{\cdot}_{sf},Q)$ gives the equation of motion in the theory of the scalar field $\varphi$ coupled to the  gauge field $\mA$. 
The gauge symmetries are given by the formula
\begin{eqnarray}
\Phi_{\varphi,\mA}\to \Phi_{\varphi,\mA}+\frac{\epsilon}{2}(Q\rho_u +[\Phi_{\varphi,\mA},\rho_u]_h).
\end{eqnarray}} 
 {\bf Proof.} The Maurer-Cartan equation for $\Phi_{\varphi,\mA}$ is
\begin{eqnarray}\label{mct}
Q\Phi_{\varphi,\mA}+\frac{1}{2}[\Phi_{\varphi,\mA},\Phi_{\varphi,\mA}]_h+\frac{1}{3!}[\Phi_{\varphi,\mA},\Phi_{\varphi,\mA},
\Phi_{\varphi,\mA}]_h=0.
\end{eqnarray}
Let us write down all terms depending on the scalar field:
 \begin{eqnarray}
&&Q\phi_{i\varphi}=2\psi_{i\p_{\mu}\p^{\mu}\varphi-im^2 \varphi},\quad 
[\phi_{i\varphi}, \phi_{\mA}]_h=2\psi_{i2[A^{\mu},\p_{\mu}\varphi]+i[\p^{\mu}A_{\mu},\varphi ]},\nonumber\\
&&[\phi_{\mA},\phi_{\mA},\phi_u]_h=4\psi_{i[A^{\mu},[A_{\mu},\varphi]]}, \quad [\phi_{i\varphi}, \phi_{i\varphi}]_h=
-4\psi_{[\varphi,\d \varphi]},\nonumber\\
&&[\phi_{i\varphi}, \phi_{i\varphi},\phi_{\mA}]_h=-\psi_{4[\varphi,[\mA, \varphi]]}.
\end{eqnarray}
The contribution of the scalar field to the Maurer-Cartan equation \rf{mct} is of the following form:
\begin{eqnarray}
&&Q\phi_{i\varphi}+[\phi_{i\varphi},\phi_{\mA}]_h+\frac{1}{2}[\phi_{\mA},\phi_{\mA},\phi_{i\varphi}]_h=0,\\
&&Q\phi_{\mA}+\frac{1}{2}[\phi_{\mA},\phi_{\mA}]_h+\frac{1}{3!}[\phi_{\mA},\phi_{\mA},\phi_{\mA}]_h+\frac{1}{2}
([\phi_{\mA},\phi_{i\varphi},\phi_{i\varphi}]_h+[\phi_{i\varphi},\phi_{i\varphi}]_h)=0.\nonumber
\end{eqnarray}
Therefore, the field equations we get are:
\begin{eqnarray}
&&\p^{\mu}\p_{\mu}\varphi-m^2\varphi+2[A^{\mu},\p_{\mu}\varphi]+[\p^{\mu}A_{\mu},\varphi]+[A^{\mu},[A_{\mu},\varphi]]=0,\nonumber\\
&&\nabla_{\mu}F^{\mu\nu}-[\varphi,\p^{\nu}\varphi]+[\varphi,[A^{\nu},\varphi]]=0.
\end{eqnarray}
We find that these equations form a system of the classical equations of motion for the action:
\begin{eqnarray}\label{sfym}
&&S_{YMsf}[\varphi,\mA]=\nonumber\\
&&\int \ud^D x((\nabla_{\mu}\varphi(x)\nabla^{\mu}\varphi(x))_K+m^2 (\varphi(x),\varphi(x))_K+\frac{1}{2}(F_{\mu\nu},F^{\mu\nu})_K),
\end{eqnarray}
where $\nabla_{\mu}\varphi=\p_{\mu}\varphi+[A_{\mu},\varphi]$. Thus the proposition is proved. $\blacksquare$\\

\noindent In order to rewrite the action for the scalar field coupled to the Yang-Mills field, one needs to define the multilinear forms 
$\{\cdot,...,\cdot\}_h$ on the complex $(\mF^{\cdot}_{sf},Q)$ like we did it in the pure Yang-Mills case. We define them in the same way, 
using the inner product \rf{sfip}. 
Moreover, we have the following proposition.

\vspace{3mm}

\noindent {\bf Proposition 4.3.} {\it The multilinear products $\{\cdot,...,\cdot\}_h$ are graded antisymmetric w.r.t. 
the ghost number.}\\
{\bf Proof.} We know that these multilinear forms are graded antisymmetric in the case of the pure 
Yang-Mills theory, i.e. for the complex 
$(\mF_{\mg}^{\cdot},Q)$. Therefore, we need to check that they are graded antisymmetric when one or more arguments belong to $(K_{\mg}^{\cdot},\t Q)$. 
For the bilinear form, this is trivial, since it is equivalent to the $Q$-invariance of the inner product \rf{sfip}. 
Now, let us consider the case of the trilinear form. The only nontrivial relation we need to check is:
\begin{eqnarray}
&&\{ \phi_u,\phi_{\mA},\phi_v\}_h=\langle [\phi_u,\phi_{\mA}]_h,\phi_v\rangle=\nonumber\\
&&\langle [\phi_u,\phi_{v}]_h,\phi_{\mA}\rangle=\{\phi_u,\phi_{v},\phi_{\mA}\}_h
\end{eqnarray}
and the proof is straightforward:
\begin{eqnarray}
&&\langle [\phi_u,\phi_{v}]_h,\phi_{\mA}\rangle=2\langle \psi_{[u,\d v]+[v,\d u]},\phi_{\mA}\rangle=\nonumber\\
&&2\int d^D x(([u(x),\p_{\mu}v(x)],A^{\mu}(x))_K+([v(x),\p_{\mu}u(x)],A^{\mu}(x))_K)=\nonumber\\
&&2\int d^D x((v(x),2[\p_{\mu}u(x),A^{\mu}(x)]+[u(x),\p_{\mu}A^{\mu}(x)])_K)=\nonumber\\
&&\langle [\phi_u,\phi_{\mA}]_h,\phi_v\rangle.
\end{eqnarray}
All other relations for the trilinear operation are just a simple consequence of the basic property of the  invariant form on the reductive 
Lie algebra $\mg$. The last step is to check the graded antisymmetry of the quadrilinear form, and the only nontrivial case that we 
need to check is:
\begin{eqnarray}
&&\{\phi_{\mA},\phi_u,\phi_v,\phi_{\mB}\}_h=\langle [\phi_{\mA},\phi_u,\phi_v]_h,\phi_{\mB}\rangle=\nonumber\\
&&\langle [\phi_{\mA},\phi_{\mB},\phi_v]_h,\phi_{u}\rangle= \{\phi_{\mA},\phi_{\mB},\phi_v,\phi_{u}\}_h.
\end{eqnarray}
The proof is as follows: 
\begin{eqnarray}
&&\langle [\phi_{\mA},\phi_u,\phi_v]_h,\phi_{\mB}\rangle=-2\int d^D x(([u(x),[v(x),A^{\mu}(x)]],B_{\mu}(x))_K+\nonumber\\
&&([v(x),[u(x),A^{\mu}(x)]],B_{\mu}(x))_K)=\nonumber\\
&&-2\int d^D x((u(x),[[v(x),A^{\mu}(x)],B_{\mu}(x)]+([[v(x),B^{\mu}(x)],A_{\mu}(x)])_K)=\nonumber\\
&&\langle [\phi_{\mA},\phi_{\mB},\phi_v]_h,\phi_{u}\rangle.
\end{eqnarray}
In this way, the proposition is proved. $\blacksquare$

\vspace{3mm}

\noindent Therefore, we can write down the homotopy Chern-Simons action for the Maurer-Cartan element (the element of ghost number 
equal to 1). 
Namely, the following proposition holds.

\vspace{3mm}

\noindent {\bf Proposition 4.4.} {\it Let $\Phi_{\varphi,\mA}=\phi_{\mA}+\phi_{i\varphi}$ be the Maurer-Cartan element, i.e. the element of the ghost number equal to 1. 
Then the homotopy Chern-Simons 
action 
\begin{eqnarray}
&&S_{HCS}[\Phi_{\varphi,\mA}]=-\frac{1}{2}\langle\Phi_{\varphi,\mA},  Q\Phi_{\varphi,\mA}\rangle\\
&&-\frac{1}{3!}\{\Phi_{\varphi,\mA},\Phi_{\varphi,\mA},
\Phi_{\varphi,\mA}\}_h-\frac{1}{4!}\{\Phi_{\varphi,\mA},\Phi_{\varphi,\mA},\Phi_{\varphi,\mA},\Phi_{\varphi,\mA}\}_h\nonumber
\end{eqnarray}
is the action of the scalar field $\varphi$ interacting with the gauge field $\mA$\rf{sfym}.}  

\vspace{3mm}

\noindent{\bf Remark.} 
Now, we remind the reader that in order to obtain the BV Yang-Mills action, 
we considered the complex $(\mH_{\mg},Q)$ which was the tensor product of 
$(\mF_{\mg},Q)$ with an integer-graded Grassmann algebra. Then, substituting the Maurer-Cartan element of the resulting complex in the homotopy Chern-Simons action (HCS), one obtains the action 
of the BV Yang-Mills theory. In the same way, one obtains 
the BV action for the theory of the scalar field coupled to the gauge field. One should consider the tensor product of the complex $(\mF_{sf}^{\cdot},Q)$ with the integer-graded Grassmann algebra and consider the Maurer-Cartan element which would be of the form:
\begin{eqnarray}
\Phi_{sf}=\rho_{\omega}+\phi_{i\varphi}+\phi_{\mA}-\psi_{i\varphi^*}-
\psi_{\mA^*}-1/2\chi_{\omega^*},
\end{eqnarray}
 where the antifield of a scalar field $\varphi^*(x)$ is the field of ghost number equal to $-1$. Then the corresponding BV 
action is the HCS action constructed by such an element:
\begin{eqnarray}
&&S_{HCS}[\Phi_{sf}]=S_{YMsf}[\varphi,\mA]+2\int d^D x (\p_{\mu}\omega(x)+[A_{\mu}(x)\omega(x)],A^{*\mu}(x))_K-\nonumber\\
&&([\omega(x),\omega(x)],\omega^*(x))_K+([\omega(x),\varphi(x)],\varphi^*(x))_K).
\end{eqnarray}

\section{The Dirac Fermion}
In this section, we consider another extension of the Yang-Mills homotopy algebra which is related to 
gauge fields coupled with the Dirac fermion. For simplicity, we consider the 4-dimensional case and 
follow the conventions about spinor fields from \cite{weinberg}. The reductive Lie algebra $\mg$ is assumed to be 
compact throughout this section. \\
{\bf 5.1. The extension of the Yang-Mills complex by the Dirac fermion.} The corresponding extension of the complex is quite similar to the one we considered in section 4:
\[
\xymatrixcolsep{20pt}
\xymatrixrowsep{0pt}
\xymatrix{
0\ar[r]&\Omega^0_{\mathfrak{g}} \ar[r]^{\ud} & \Omega^1_{\mathfrak{g}} \ar[r]^{\m}  &\Omega^1_{\mathfrak{g}} \ar[r]^{\di} & \Omega^0_{\mathfrak{g}}\ar[r]&0\\
 && \bigoplus & \bigoplus     &&\\
 &0\ar[r]& \mathcal{S}_{\mathfrak{g}}\ar[r]^{D}   & \mathcal{S}_{\mathfrak{g}}\ar[r]&0,
}
\]
where $\mathcal{S}_{\mathfrak{g}}=\mathcal{S}\otimes \mathfrak{g}$, $\mathcal{S}$ is the space 
of 4-dimensional Dirac spinors and  $D=\gamma^{\mu}\p_{\mu}+m$ is the Dirac operator. The realization 
of this complex in a superspace is as follows. Let us consider the following objects: 
\begin{eqnarray}
\phi_{\xi}=e_1\xi(x),\quad \psi_{\eta}=e_2\eta(x),
\end{eqnarray}
which are associated with the Dirac spinor fields $\xi(x)$ and $\eta(x)$ of fermion statistics. 
Here $e_1$ is even and $e_2$ is odd. We consider the spaces $\mD_{\mg}^1$ and $\mD_{\mg}^2$ 
spanned by the elements of type $\phi_{\xi}$ and $\psi_{\eta}$ respectively. One can construct the differential operator related to the Dirac one $Q'=2e_2\frac{\p}{\p e_1}D$. We get a chain complex $(\mD^{\cdot}_{\mg},Q')$: $0\to \mD_{\mg}^1 \xrightarrow {Q'}\mD_{\mg}^2\to 0, \quad Q'\phi_{\xi}=2\psi_{D\xi}$, where the grading operator is given by the expression $N'_g=e_1\frac{\p}{\p e_1}+2e_2\frac{\p}{\p e_2}$. One can consider the chain complex 
\begin{eqnarray}\label{dfymcomplex}
0\to\mg\xrightarrow{i}\mathcal{F}_{Df}^{0}\xrightarrow{Q^{tot}}\mathcal{F}_{Df}^{1}
\xrightarrow{Q^{tot}}\mathcal{F}_{Df}^{2}\xrightarrow{Q^{tot}}\mathcal{F}_{Df}^{3}\to 0
\end{eqnarray}
such that 
\begin{eqnarray}\label{dfsum}
&&\mathcal{F}_{Df}^{i}=\mathcal{F}_{\mg}^{i}\oplus \mathcal{D}_{\mg}^{i} \quad (i=1,2),\nonumber\\  
&&\mathcal{F}_{Df}^{j}=\mathcal{F}_{\mg}^{j} \quad (j=0,3)
\end{eqnarray}
and $Q^{tot}=Q+Q'$. As in the case of the scalar field, in this section we will denote $Q^{tot}$ as $Q$, the total grading operator $N^{tot}_g=N_g+N'_g$ and the complex \rf{dfymcomplex} as $(\mathcal{F}^{\cdot}_{Df},Q)$ . The $Q$-invariant inner product  
(in the sense of Proposition 2.1.) on the complex $(\mF_{Df}^{\cdot},Q)$ can be defined in the following way. 
Let  
$\Phi=\rho_{u}+\phi_{\mathbf{A}}+\phi_{\xi}+ \psi_{\mathbf{W}}+\psi_{\eta}+\chi_{v}$ and 
$\Phi'=\rho_{u'}+\phi_{\mathbf{A}'}+\phi_{\xi'}+ \psi_{\mathbf{W}'}+\psi_{\eta'}+\chi_{v'}$, the pairing is given by:
\begin{eqnarray}\label{dfip}
&&\langle\Phi,\Phi'\rangle=\int d^4x((\mA,\mathbf{W}')(x)+(\mW,\mA')(x)-\nonumber\\
&&2(u(x),v'(x))_K-2(v(x),u'(x))_K+(\b \xi(x),\eta'(x))_K+(\b \eta'(x),\xi(x))_K+\nonumber\\
&&(\b \xi'(x),\eta(x))_K+(\b \eta(x),\xi'(x))_K
\end{eqnarray} 
such that the direct sum in \rf{dfsum} is orthogonal. 

\vspace{3mm}

\noindent{\bf 5.2. The $L_{\infty}$ algebra on the complex $(\mF^{\cdot}_{Df},Q)$.} In this subsection, we will give explicit formulas for the extension of the Yang-Mills $L_{\infty}$ algebra which will lead to the Lagrangian of the gauge field coupled to a Dirac fermion. First, we give the definition of the modified bilinear operation.

\vspace{3mm}

\noindent {\bf Definition 5.1.} {\it The bilinear operation on the complex $(\mF^{\cdot}_{Df},Q)$ 
\begin{eqnarray}
[\cdot, \cdot]_h: \mathcal{F}^i_{Df}\otimes \mathcal{F}^j_{Df}\to \mathcal{F}^{i+j}_{Df}
\end{eqnarray}
is defined by the relations \rf{ymha} and, in the case when one of the arguments belongs to the subcomplex $(\mD_{\mg}^{\cdot},Q')$, by 
\begin{eqnarray}
&&[\phi_{\mA},\phi_{\xi}]=2\psi_{[\hat \mA, \xi]}, \quad [\phi_{\xi},\phi_{\eta}]=-2\psi_{(\xi,\eta)}, 
\quad [\phi_{\xi},\psi_{\eta}]=\chi_{[\b \eta,\xi]-[\b \xi,\eta]},\\
&&[\phi_{\mA},\psi_{\xi}]=0,\quad [\psi_{\mW},\phi_{\xi}]=0,\quad [\rho_u,\phi_{\xi}]=2\phi_{[u,\xi]}, 
\quad [\rho_u,\psi_{\xi}]=2\psi_{[u,\xi]},\nonumber
\end{eqnarray} 
where $(\xi,\eta)^{\mu}=[\b \xi,\gamma^{\mu}\eta]+[\b \eta,\gamma^{\mu}\xi]$ and $\hat\mA=\gamma^{\mu}A_{\mu}$. }

\vspace{3mm} 

\noindent {\bf Remark.}
In contrast to the previous case of a scalar field, there is no need to modify the trilinear 
operation (it is related to the fact that the corresponding nonlinear field equation contains fermions only 
in the terms of the second order in fields (see below). 

\vspace{3mm} 

\noindent The following proposition holds.

\vspace{3mm} 

\noindent {\bf Proposition 5.1.} {\it The bilinear operation from Definition 5.1. and the trilinear operation 
defined by \rf{ymha} satisfy the relations of the $L_{\infty}$ algebra \rf{rel}}.\\
{\bf Proof.} First, we prove that $Q$ acts as a derivation on the bracket $[\cdot,\cdot]_h$, namely 
\begin{eqnarray}
Q[a_1,a_2]_h=[Q a_1,a_2]_h+(-1)^{n_{a_1}}[a_1,Q a_2]_h.
\end{eqnarray} 
Let us take $a_1=\rho_u$, $a_2=\phi_{\xi}$. Then
\begin{eqnarray}
&&[Q\rho_u,\phi_{\xi}]_h+[\rho_u,Q\phi_{\xi}]_h=2[\rho_{\d u},\phi_{\xi}]_h+2[\rho_u,\phi_{D\xi}]_h=\nonumber\\
&&4\psi_{D[u,\xi]}=Q[\rho_u,\phi_{\xi}]_h.
\end{eqnarray}
Another nontrivial case is when $a_1=\phi_{\xi}$, $a_2=\phi_{\eta}$:
\begin{eqnarray}
&&Q[\phi_{\xi},\phi_{\eta}]_h=2\chi_{\p_{\mu}([\b \xi,\gamma^{\mu}\eta]+[\b \eta, \gamma^{\mu}\xi])}=\nonumber\\
&&2\chi_{[\b \eta,D\xi]-[\overline{D\xi},\eta]}+\chi_{[\b \xi,D\eta]-[\overline{D\eta},\xi]}=
[Q\phi_{\xi},\phi_{\eta}]-[\phi_{\xi},Q\phi_{\eta}].
\end{eqnarray} 
Next, we check the homotopy Jacobi identity in the case when one of the arguments belongs to $\mD_{\mg}$. In this 
case, the homotopy Jacobi identity reduces to the usual one: 
\begin{eqnarray}\label{jacdf}
[a_1,[a_2, a_3]_h]_h-[[a_1,a_2]_h, a_3]_h-(-1)^{n_{a_1}n_{a_2}}[a_2,[a_1, a_3]_h]_h=0.
\end{eqnarray}
The only nontrivial case here is when $a_1=\phi_{\mA}$, $a_2=\phi_{\xi}$, $a_3=\phi_{\eta}$. Let us  write down each term 
separately:
\begin{eqnarray}
&&[\phi_{\mA},[\phi_{\xi},\phi_{\eta}]_h]_h=2\chi_{[A_{\mu},[\b \xi,\gamma^{\mu}\eta]+[\b \eta,\gamma^{\mu}\xi]]},\nonumber\\
&&[[\phi_{\mA},\phi_{\xi}]_h,\phi_{\eta}]_h=2[\psi_{[\hat\mA,\xi]},\phi_{\eta}]=2\chi_{[\b \eta,[\hat \mA,\xi]]+
[[A_{\mu},\xi],\gamma^{\mu}\eta]},\nonumber\\
&&-[\phi_{\xi},[\phi_{\mA},\phi_{\eta}]_h]_h=2\chi_{[\b \xi,[\hat \mA,\eta]]+[[A_{\mu},\eta],\gamma^{\mu}\xi]}.
\end{eqnarray}
Summing all terms, we find that the Jacobi 
relation is satisfied. Thus, the proposition is proved. $\blacksquare$ 

\vspace{3mm}

\noindent{\bf 5.3. The generalized Maurer-Cartan, a homotopy Chern-Simons and the Dirac fermion coupled to the gauge field.} 
As in section 4, we write down the Generalized Maurer-Cartan equation in the case of the 
homotopy Lie algebra, we obtained in the previous subsection, and provide the physical interpretation. 
Namely, the following proposition holds.

\vspace{3mm}

\noindent{\bf Proposition 5.2.} {\it Consider the Maurer-Cartan element $\Phi_{\xi,\mA}=\phi_{\xi}+\phi_{\mA}$ (the element of ghost number equal to 1) in the complex $(\mF_{Df}^{\cdot},Q)$. 
The Generalized Maurer-Cartan equation for $\Phi_{\xi,\mA}$ gives the equation of motion in the 
theory of a Dirac fermion field $\xi$ coupled with a gauge field $\mA$, and the gauge symmetries 
are given by the formula
\begin{eqnarray}
\Phi_{\xi,\mA}\to \Phi_{\xi,\mA}+\frac{\epsilon}{2}(Q\rho_u +[\Phi_{\xi,\mA},\rho_u]_h).
\end{eqnarray}}
{\bf Proof.} The Maurer-Cartan equation 
\begin{eqnarray}
Q\Phi_{\xi,\mA}+\frac{1}{2}[\Phi_{\xi,\mA},\Phi_{\xi,\mA}]_h+\frac{1}{3!}[\Phi_{\xi,\mA},\Phi_{\xi,\mA}, \Phi_{\xi,\mA}]_h=0
\end{eqnarray}
decomposes into two parts:
\begin{eqnarray}\label{ymrhs}
&&Q\phi_{\mA}+\frac{1}{2}[\phi_{\mA},\phi_{\mA}]_h+\frac{1}{3!}[\phi_{\mA},\phi_{\mA},\phi_{\mA}]_h+
\frac{1}{2}[\phi_{\xi},\phi_{\xi}]_h=0,\nonumber\\
&&Q\phi_{\xi}+[\phi_{\mA},\phi_{\xi}]_h=0.
\end{eqnarray}
It is easy to check that equations \rf{ymrhs} coincide with
\begin{eqnarray}
\nabla_{\mu}F^{\mu\nu}=\b \xi\gamma^{\nu}\xi, \quad (\gamma^{\mu}\p_{\mu}+m)\xi+[\hat\mA,\xi]=0,
\end{eqnarray}
which precisely coincide with the equations of motion given by the action of the Dirac fermion 
coupled to the gauge field:
\begin{eqnarray}\label{fgf}
S_{YMdf}=\int d^4 x (\frac{1}{2}(F_{\mu\nu}, F^{\mu\nu})_K-2(\b\xi,(\gamma^{\mu}\p_{\mu}+m)\xi)_K-2(\b\xi,[\hat\mA,\xi])_K.
\end{eqnarray}
Thus the proposition is proved. $\blacksquare$

\vspace{3mm}
 
\noindent In order to represent the action \rf{fgf} in the homotopy Chern-Simons form, one 
has to generalize the 
multilinear forms $\{\cdot,..., \cdot \}_h$ to our case. It is easy to see that these forms are also graded 
antisymmetric with respect to the ghost number. Therefore, one can write the 
HCS action for the Maurer-Cartan element $\Phi_{\xi,\mA}$ and it is not hard to check that it coincides with \rf{fgf}. 

As in the case of the scalar field, one can introduce the antifield for the Dirac fermion by means of a tensor product of the complex 
$(F_{Df},Q)$ with the integer-graded Grassmann algebra. Let us consider a Maurer-Cartan element 
(now this is an element of $total$ ghost number equal to 1) and represent it in the form: 
$\Psi_{Df}=\rho_{\omega}+\phi_{\xi}+\phi_{\mA}-\psi_{\xi^*}-
\psi_{\mA^*}-1/2\chi_{\omega^*}$ ($\xi^*$ has the sense of the antifield). If one substitutes 
$\Psi_{df}$ into the HCS action, it is easy to see that it coincides with the BV action for the gauge field coupled to the Dirac fermion:
\begin{eqnarray}
&&S_{HCS}[\Phi_{df}]=S_{YMdf}+2\int d^4 x (\p_{\mu}\omega(x)+[A_{\mu}(x)\omega(x)],A^{*\mu}(x))_K-\\
&&([\omega(x),\omega(x)],\omega^*(x))_K+2([\omega(x),\b \xi^*(x)],\xi(x))_K+2([\omega(x), \b\xi(x)],\xi^*(x))_K)\nonumber.
\end{eqnarray}

\section{Final Remarks}

In this paper, we considered the basic models of the Field Theory related to the nonabelian gauge theory, namely the models of fermion and scalar fields interacting  with Yang-Mills fields. 
We have shown that as it was in the case of the pure Yang-Mills theory, 
these actions can be represented in the homotopy Chern-Simons form, such that the equations of motion are the generalized 
Maurer-Cartan equations for some homotopy Lie algebra. Moreover, the gauge transformations of these theories naturally emerged from the symmetry transformations of these Maurer-Cartan equations. 

The generalization to noncommutative field theories \cite{sw} is quite straightforward:  
it is not hard to see that the noncommutative gauge theories can be expressed in the homotopy Chern-Simons 
form if one modifies a little bit the algebraic operations by means of the Moyal star. However,  the gauge group should be 
$U(N)$ due to the no-go theorem \cite{sw}.

We also mention some attempts to obtain the (BV) YM action from the open SFT \cite{taylor}-\cite{feng}. The connection with our formalism is not yet clear. 

Another important question is related to Gravity. The Zwiebach 
formulation of the closed String Field Theory via the $L_{\infty}$ algebra on the closed string states suggests that there should be a formulation of the Einstein-Hilbert action (together with the Kalb-Ramond field and the dilaton) in terms of some homotopy Lie algebra on the space of fields. The problem is that in order 
to do this, one needs to expand the metric field $G_{\mu\nu}$ around some flat metric $\eta_{\mu\nu}$ by means of some formal 
parameter: $G_{\mu\nu}=\eta_{\mu\nu}+th_{\mu\nu}+t^2s_{\mu\nu}+...$, leading to the 
background dependence (due to nonuniqueness of such a choice of flat metric and formal parameter) of this construction. 
Nevertheless, in \cite{zeit3}, we made a few steps in this direction. The problem is that the structure of Gravity in this approach is much more complicated than that of the Yang-Mills and related field theories: the structure of operators at the second order, 
namely their relation to original expansion of metric field is quite mysterious. There is a $bivertex$ 
operator (the corresponding ``bivertex'' field is $h_{\mu\rho}\eta^{\rho\sigma}h_{\sigma\nu}$), 
depending on the first order of expansion of the metric field. In papers \cite{zeit}, 
\cite{zeit3}, we have found an explanation by using the conjectured relation to the conformal invariance 
condition of the sigma-model, however, there is no canonical way to continue this construction to all orders. 
Moreover, in these papers we found that in this approach the diffeomorphism symmetries of Gravity would have an $algebroid$-like structure. 

Another approach to Gravity can be related to the first order formulation of the string theory from $\cite{lmz}$, $\cite{zeit2}$. There, we introduced the twistor-like variables for the metric field, B-field and dilaton. Then the symmetries are reduced to the holomorphic ones. We expect that this formulation can be very helpful with respect to our approach, since the first order sigma-model theory becomes free of some difficulties which are present in the usual one: 
there are no contact terms and the perturbation theory does not destroy the geometrical setting.

We also mention that it will be interesting to find the relation of our formalism with what is known as 
$the$ $unfolded$ $dynamics$ approach, also leading to the $L_{\infty}$ algebras, related to the geometric structures associated with higher spin theories (see \cite{bekaert}, \cite{vas} and furher references therein). 

We will consider a part of problems sketched above in the forthcoming papers.

\section*{Acknowledgements} 
I would like to thank D. Borisov, I.B. Frenkel, M.M. Kapranov, M. Movshev, T. Pantev, M. Rocek, H. Sati, J. Stasheff, D. Sullivan, M.A. Vasiliev and G. Zuckerman for numerous discussions on the subject 
and also I.B. Frenkel and N.Yu. Reshetikhin for their permanent encouragement and support. I am grateful to R. Cheung for reading a preliminary version of this paper and to the Referee for his attention and a lot of important comments.

\end{document}